\documentstyle[preprint,aps]{revtex}
\newcommand{\be}{\begin{equation}}
\newcommand{\ee}{\end{equation}}
\newcommand{\ba}{\begin{eqnarray}}
\newcommand{\ea}{\end{eqnarray}}
\begin{document}
\draft
\title{Mean Field Theory of the Morphology Transition in Stochastic Diffusion
Limited Growth}
\author{Yuhai Tu}
\address{IBM T. J. Watson Research Center, P. O. Box 218, Yorktown Heights,
NY 10598}
\author{Herbert Levine}
\address{Department of Physics and
Institute for Nonlinear Science
University of California, San Diego
La Jolla, CA  92093-0402}

\maketitle

\begin{abstract}

We propose a mean-field model for describing the averaged properties
of a class of stochastic diffusion-limited growth systems. We then show
that this model exhibits a morphology transition from a dense-branching
structure with a convex envelope to a dendritic one with an overall concave
morphology.
We have also constructed an order parameter
which describes the transition quantitatively. The transition is shown to be
continuous, which can be verified by noting the non-existence of any
hysteresis.

\end{abstract}
\pacs{68.70.+w,47.15.Hg,47.20.Hw}

\newpage
\section{Introduction}

Diffusion limited growth processes occur in a broad range of interesting
systems
ranging from physics to chemistry and to biology~\cite{a1}.
The common features
that we observe in these systems are due to the patterns being driven
by a generic instability which
arises when the process of one phase replacing another phase is controlled
by a diffusive field~\cite{a2}. This instability leads to the breakdown of any
simple shape and thereby causes the formation of complex interfacial
structures.  The structures that do form vary from smooth fingers
to dendritic arbors to disordered, possibly fractal patterns in a manner
dependent on the details of different individual systems,

Most of the work in this field has utilized one of two approaches. The first
method relies on a continuum description of the interface evolution
process, leading to a free surface problem for partial differential equations
governing the diffusive transport.
The models that one arrives at are purely deterministic; stochastic
behavior can of course arise dynamically due to the inherent nonlinearities
in the interfacial dynamics.
A major success of this line of reasoning has been the
discovery of the ``microscopic-solvability" criterion, which
enables us to understand the selection of a unique finger pattern out of
an apparently continuous family of available solutions~\cite{a1}.
More recently~\cite{a2'}, this approach has been used to study the global
morphology
of diffusion-limited solidification. From our perspective, studies of
solidification using the phase-field method~\cite{a2''} fall
into this same category;
although quite different in computational detail from the sharp
interface equations, these methods also embody
deterministic microscopic models for interfacial evolution.

The second general scheme available to investigate diffusion-limited
growth is the kinetic approach, where the diffusion field is represented
by particles executing random walks. The
boundary condition for the diffusive field and the local dynamics of the
interface are combined into microscopic rules for these random walkers to stick
to the aggregate cluster representing the growing ``solid" phase.
The simplest member of
this class of models is diffusion limited aggregation (DLA) ~\cite{a3}, where
the
sticking rule is simply that the walker will become part of the aggregate
upon contact with the aggregate. This
simple model is known to produce fractal structure and has been
extensively investigated ~\cite{a4}. However, it is fair to say that
a full theory of the relationship between the
diffusive nature of the controlling field (walkers) and the fractal structure
of
the cluster which DLA generates still eludes us. Note that models of
this second type are explicitly stochastic; whether this represents
an important consideration or not is still an open question.

In order to study more realistic solidification processes,
several groups have introduced more complicated local
kinetics(sticking rules) and in addition have used
a finite density of random walkers to mimic finite undercooling effects..
Saito and Ueta~\cite{a5}
performed simulations of a many-walkers  model with  Ising-like sticking
probabilities. Liu and Goldenfeld~\cite{a6} used a relaxation method in
updating the walker distribution function (instead of using walkers per se)
but chose a sticking
rule which does not have any well defined thermodynamic limit.
In a series of recent papers,
Shochet et al ~\cite{a7,a8} have introduced
the ``diffusion-transition scheme", which
they have combined these above two ideas, i.e. directly
solving for the walker distribution and using an Ising-like sticking
probability at the interface. The main purpose of this study was to investigate
the existence of different patterns as one changes the various control
parameters, such as the density of the diffusive field at the boundary of
the system(
the undercooling), the chemical potential difference between the two phases,
(in a spin system analogy, this corresponds to a magnetic field
favoring the solid phase),
and the surface energy (bond energy, in the spin analogy).

In all the studies mentioned above, a morphology transition
from a dense branching
morphology (DBM) to a dendritic structure has been observed.
Locally, the DBM phase
resembles the ramified structure of a DLA fractal, but at
larger length scale, it is densely packed, i.e. it has a finite density
and the pattern has a well-defined smooth envelop. For the dendritic
phase, the directions of
the dendrites are determined by the anisotropy of either the surface energy or
the interfacial kinetic coefficient.
Due to the probabilistic nature of these models, each
realization of the pattern looks different; in order to characterize
the morphology transition, an ensemble average is used to
study the statistical properties of the pattern distribution function.
In terms of this ensemble averaged pattern, the DBM to dendrite transition is
connected to the change of the envelop shape from convex to concave.

Even though such a transition is quite well established in numerical
simulations, there has been very few theoretical attempts to understand
the transition~\cite{a9}. Such a study is quite crucial if we are to answer
questions
regarding the role of explicit stochasticity, the possibility of
morphology selection via a maximum velocity selection
principle~\cite{a13} and in fact, the whole concept of a sharp transition
between distinct
morphologies.

Our goal here is to propose a new mean field theory (MFT) which
describes the ensemble
averaged behavior of these stochastic models, and more specifically
the morphology transition. A mean field is again a deterministic
set of evolution equations, but these are meant to describe the
average of the overall probability distribution and not any given
realization; they should not be confused with the deterministic
microscopic models mentioned above. In the next section, we describe our
previous work on mean field theories for the DLA sticking rule and how
one can phenomenologically introduce mean field reaction rates which
lead to a better (i.e. more physically motivated) set of equations. Afterwards,
we explicitly discuss the issue of the nature of the morphology transition
and conclude that in the mean field treatment, it is continuous. The final
section contains a concluding discussion.

\section{Mean Field Equations}

In our previous work~\cite{a10}, we have studied the existence and
nature of a morphology transition
for the following set of equations:
\ba
\dot{\rho}&=&u(\rho ^{\gamma}+a^2\nabla ^2\rho)\\
\dot{u}&=&D\nabla^2 u-\dot{\rho}
\ea
where $D$ is the diffusion constant, $\rho$
is the density of the cluster, $u$ is the density of the walkers,
with the boundary conditions $u(\infty)=\Delta$, $\rho(\infty)=0$.
Eq. (2) arises due to the conservation of particles and the
diffusive nature of the random walkers. Eq.(1) is determined by the
local kinetics of the growth process.
The above eqs. (1,2) were originally proposed to describe ensemble averaged
behavior of DLA without the $\dot{u}$ term in the second equation, and with
different
boundary condition for $u(\infty)$~\cite{a11}. The crucial
difference between these equations and the original Witten-Sander DLA
mean field theory is that the phenomenological parameter $\gamma$
is taken to be strictly greater than unity; later~\cite{a10'}, this type of
cutoff in the growth rate at small density was derived by proper
inclusion of the average effects of the multiplicative noise that
must be added to the naive reaction kinetics term so as
to make a complete DLA theory.  In that study ~\cite{a10}, we
did indeed discover that as one lowers the undercooling $\Delta$,
there is a morphology transition characterized by the change of
the envelop shape from convex(DBM) to concave(dendrite). We did not
find any discontinuity in the velocity slop versus $\Delta$; this was
subsequently verified to be true of the actual transition in
many-walkers DLA ~\cite{a10''}.

It is obvious from eqs. (1,2) that they cannot describe the kinetics
of the more complex simulations; since the model was supposed to
mimic the sticking rules of pure DLA, there is no parameter in
the mean field model playing the role of the chemical potential
difference between the two phases. Finding an augmented set of equations
is important because the aforementioned results on the continuous
nature of the transition appear to disagree with the findings
of the Shochet et al diffusion-transition simulation. Also, a more
physical model would obviate the need for the parameter $\gamma$, since
as we shall see, a cutoff in growth at low density is an immediate
consequence of the actual kinetics of these more realistic simulations.

We now turn to construction of the new mean field model.
One way of understanding the crucial feature which we need
to include is by recognizing that the local
kinetics satisfy detailed balance. The local energy functional
which governs this balance is determined
by both the surface energy and the chemical potential difference between
the two phases. Thus, both the solidification of
the liquid and the melting of the solid occur simultaneously
with their rates dependent on this local energy functional. This
leads immediately to the form of the simplest MFT
as:
\ba
\dot{\rho}&=&(1-\rho)R_{s}(\Delta E,\mu_{s},u)-\rho R_{m}(\Delta E,\mu_{s},u)\\
\dot{u}&=&D\nabla^2 u-\dot{\rho}
\ea
The first term on the right hand side of eq. (3) represents the product of the
probability of the site being empty $(1-\rho)$ and the rate of solidification
$R_{s}(\Delta E,\mu_s,u)$; similarly, the second represents the product of the
probability of the site being occupied by solid $\rho$ and the rate of melting
$R_{m}(\Delta E, \mu_{s},u)$. Here
$\Delta E$ is the local surface energy and
$\mu_s$ is the chemical potential difference between the two phases.

In the actual kinetic simulation, the surface energy can depend on the detailed
geometry of the solid near the vicinity of a possible site for melting or
solidifying. In the mean field description,
$\Delta E$ at site $i$ will be taken to depend on $\bar{\rho}_{i
}$, which is the average of all nearest-neighbor densities at site $i$. In
the continuum limit, $\bar{\rho}=\rho+(a^2/2d)\nabla^2\rho$,
where $a$ is the lattice spacing and $d$ is the dimension. There is no
analytic derivation for the explicit
form of $\Delta E(\bar{\rho})$; however, we know that $\Delta E\rightarrow
\infty$ as $\bar{\rho}\rightarrow 0$ in order to suppress the nucleation
process
inside the liquid phase
and $\Delta E\rightarrow -\infty$ as $\bar{\rho}\rightarrow 1$ to
suppress melting inside the solid. Also $\Delta E=-2B,0,2B$ when $\bar{\rho}
=\frac{1}{4},\frac{1}{2},\frac{3}{4}$, which reflects the fact that when
a site has 1, 2 or 3 neighbors being occupied, the bond energy gain
energy for this site to be
occupied is 2B, 0 or -2B. We have chosen an expression for $\Delta E$ which
satisfies the above requirements:
\be
\Delta E(\bar{\rho_{i}})=2B/tg(\pi\bar{\rho_{i}})
\ee
We stress that the explicit form of $\Delta E(\bar{\rho_{i}})$ is
chosen for convenience, once we ensure that it has sensible limits. It will
become clear later in the paper that the detailed form of $\Delta E$ is
unimportant for the qualitative behavior of the system,
which in any event is all MFT can offer. Once we have an expression for
$\Delta E$,
the simplest choices for the transition rates are:
\be (A)\;\;\;\;R_{s}(\Delta E,\mu_{s},u)=\frac{u}{1+\exp(\Delta E-\mu_{s})}\;\;
;\;\;R_{m}(\Delta E,\mu_{s},u)=\frac{1}{1+\exp(-\Delta E+\mu_{s})}
\ee
\be (B)\;\;\;\;R_{s}(\Delta E,\mu_{s},u)=\frac{u}{u+\exp(\Delta E-\mu_{s})}\;\;
;\;\;R_{m}(\Delta E,\mu_{s},u)=\frac{1}{1+u \exp(-\Delta E+\mu_{s})}
\ee
in accordance with the algorithms used by references ~\cite{a7} and ~\cite{a5}
respectively.
We can easily see that the ratio of the two transition rate are the same
for the two algorithms; this is what matters for the morphology
transition, as we
will show later. For the rest of the paper, we will use scheme (A).

\section{Results}

We have studied the above equations numerically for both one and
two dimensions. We first present our results in 1D. There are three
variables that are important for the morphology transition, i. e., $B$,
$\mu_s$ and $\Delta$. We choose to fix $B=0.7$, $\Delta=0.7$ and vary
the chemical potential $\mu_s$. We find that there is a critical value
$\mu^{*}_{s}(B,\Delta)$, such that when $\mu_s >\mu^{*}_{s}$
the dynamics approaches a steady state. That is, starting form
any initial condition, the system settles into a state in which the solid
phase replaces the liquid phase with a constant velocity  and the
profiles of the $\rho$ and $u$ fields are time independent in the co-moving
frame. When $\mu_s <\mu^{*}_{s}$, there is no steady state solution. Instead,
the $\rho$ field will grow up to the maximum density $\rho=1$.
and the profile of the liquid density
is now time-dependent, and the width of the $u$ field
increases with time.
These two phases of eqs. (3,4) are shown in figure 1(a), (b).

Before describing our results for the more interesting 2D case, let us
try to understand the above results. We
look for steady state solution of eq.(3,4), where the interface (or the
front) moves with certain velocity $v$, and shape of the field profile
does not change with time. We transform to the
co-moving frame by the change of variable: $z=x-vt$; the equations (3,4)
become:
\ba
-v\partial\rho/\partial z&=&\frac{u(1-\rho)}{1+\exp (\Delta
E-\mu_s)}-\frac{\rho}{1+\exp (-\Delta E+\mu_s)}\\
-v\partial\rho/\partial z&=&D\partial^2 u/\partial z^2+v\partial v/\partial z
\ea
The second equation above can be integrated to give:
\be
\rho=\Delta-u-\frac{D}{v}\partial u/\partial z
\ee
where the boundary conditions $\rho\rightarrow 0$, $u\rightarrow \Delta$ (as
$z\rightarrow \infty$)has been used.
Next, we consider the profiles of the $\rho$ and $u$ fields away from the
front,
where there is no spatial dependence, so:
\be
\frac{u(1-\rho)-\rho\exp(\Delta E-\mu_s)}{1+\exp (\Delta E-\mu_s)}=0
\ee
\be
\rho+u=\Delta
\ee
The solution of eq.(11) is: $\rho=0$ (recall, that $\Delta E \ \rightarrow
\ \infty$ as $\rho \ \rightarrow \ 0$) or $\rho=1$ ($\Delta E \ \rightarrow \
-\infty$) or $u=\frac{\rho}{1-\rho}\exp
(\Delta E-\mu_s)$ ($0<\rho<1$). The latter nontrivial relation is
plotted in figure 2 together with the
straight line $\rho+u=\Delta$ in the $u-\rho$ plane. For $\mu_s$ large enough,
there are three fixed points A, B and C determined by the eqs. (11,12). Their
relative position is illustrated in fig. 2, it can be easily seen that fixed
points A and C are stable, whereas fixed point B is unstable and it separates
the attraction basin of the two stable fixed points. Fixed point A represents
the liquid state where $\rho=0$, $u=\Delta$, and fixed point $C$ with
$\rho\sim\Delta$ and $u<<\Delta$ describes the solid state.

It is well known in non-equilibrium systems~\cite{a12}, when a stable state
invades another stable state, the velocity of the front is uniquely determined.
One way to see
this is to substitute the solution of eq. (10) into eq. (8), we have an
equation for just $u$:
\be
D\partial^2 u/\partial z^2 +[v-\frac{D}{v}\frac{u+\exp (\Delta E-\mu_s)}{1+\exp
(\Delta E-\mu_s)}]\partial u/\partial z=\frac{u(1-\Delta +u)-(\Delta-u)\exp
(\Delta E-\mu_s)}{1+\exp (\Delta E-\mu_s)}
\ee
The front velocity $v$ is now a parameter in the above equation. By taking into
account the asymptotic behavior of the $u$ field at $z\rightarrow\pm\infty$,
one can show the above equation is an eigenvalue equation for $v$, i. e.,
it is only solvable for an unique value of $v$.

At some critical plane in $(\mu_s,\Delta,B)$
space, the fixed points B and C disappear together, which means there is no
steady state growth
possible in this regime. In fact, depending on the detailed functional form
of $\Delta E$, the steady state solution disappears before the merging of
B and C when there is no solution for the eigenvalue problem for any $v>0$.
This is the case for our choice of $\Delta E$. In the regime where there is
no steady state, the solid phase flows into the maximum density state with
$\rho=1$ ($\rho=1$ is a still a fixed point for eq.(3)), and since
the liquid phase only supplies a density of $u(\infty)
=\Delta<1$, there is a deficit in the $u$ field. The u field tries to
compensate for this deficit by getting particles from regions
increasingly deeper into the liquid side
with increasing
time and thereby develops a time dependent profile. We therefore call this
phase of the
dynamics the ``starved phase", and the previous one the ``saturated phase".

Having identified the two phases in 1D, we can proceed to study the more
important 2D case. We use the discretized version of eqs. (3,4)
using the grid size
$\Delta x=\Delta y=a$, because any structure which is smaller than the lattice
spacing
$a$ is unphysical. Let $a=1$ and by writing:
$\bar{\rho}(i,j)=\frac{1}{4}(\rho(i+1,j)
+\rho(i-1,j)+\rho(j+1,i)+\rho(j-1,i))$, the surface energy anisotropy is
automatically included in our model. The time step is chosen
to be small $\Delta t=0.01$.
We find that both of the two phases found in 1D have
corresponding states in 2D. When the chemical potential is large, there
is a 2D steady state solution in which, in analogy with
the ``saturated phase" in 1D,
the solid phase has uniform density and the contour
line separating the solid and liquid phase has a convex shape (deformed from a
circle due to anisotropy). This phase can be identified as the DBM phase.
The 2D ``starved phase" becomes much more interesting due to
the two dimensionality and the presence of anisotropy.
Because of the surface energy anisotropy, the
solid phase has four preferred growth directions
	($45^0$ with respect to the lattice)
to grow. As in the 1D case, the solid phase attempts to grows with maximum
density $\rho \ = \ 1(>\Delta)$.
However, due to the
two dimensionality, the deficiency of supply can be compensated by
the screening
effect, by which the growth of solid in the preferred directions screens the
growth in other direction, and therefore a steady state can be reached.
This possibility is directly connected to the existence of
the Mullins-Serkerka instability. In fact, a stability analysis~\cite{a14}
for our previous
model shows that there is indeed such instability for the flat interface
in the  dendritic phase.
A succession of snaphots of the morphology as we go through
the transition by varying the chemical potential is shown in fig.3 .

So far, we have shown that eqs. (3,4) have a
morphology transition in 2D between
two phases whose origin can be related to the one dimensional behavior of
eqs.(3,4). In order to better characterize
the transition, we need to have an order parameter which describes
the different macroscopic nature of the two different morphologies in
analogy to what is normally done for an
equilibrium phase transition. In ~\cite{a7}, Shochet et al proposed the
front velocity as an order parameter; this was  based on previous conjectures
concerning the selection of coexisting patterns~\cite{a13}. In their
simulation, they showed that
the dependence of the tip velocity on the chemical potential has a
discontinuity of slope at the transition point (on a log-log plot). We have
therefore
measured the tip velocity versus the chemical potential for our model and
the results are shown in figure 4(a), it is quite evident from the plot
that there is no drastic change at the transition point, in agreement with the
results of our previous model~\cite{a10}. We as yet have no explanation for
this discrepancy.

In general, a useful way to construct an
order parameter in non-equilibrium systems
is to first study the linear stability of the system near the transition.
Then the projection of the field onto the most unstable mode or the amplitude
of the most unstable mode can be used as an order parameter. This amplitude
equation approach is useful in many non-equilibrium systems, including Raleigh
Benard
convection and Taylor-Couette flow~\cite{a12}.
However, this weakly
non-linear methodology
does not apply to our case, because dendritic growth is highly nonlinear. The
initially most unstable mode strongly interacts with other unstable modes
in the system
and the scale of the final pattern is determined by the system size, not by
the length scale set by the most linearly unstable mode. We are therefore
forced to find some alternative way to construct an order parameter.

Given that the very nature of the transition is tied to the change of
the spatial pattern, it seem natural to us to pick as a measure of
the transition a quantity describing how the dendritic pattern picks up
global correlation as compared to the DBM state. Recall that in the dendritic
phase, the solid behind the front
has a highly non-uniform density; we thus define our order parameter as the
standard
deviation of the solid density:
\be
\psi=\sqrt{<\rho^2>-<\rho>^2}
\ee
with $<\rho^2>=\frac{\int_{\omega}\rho^{2}d\vec{x}}{\int_{\omega}d\vec{x}}$ and
$<\rho>=\frac{\int_{\omega}\rho d\vec{x
}}{\int_{\omega}d\vec{x}}$.
The integration region $\omega$ is the solid dominated region defined as where
$u<\Delta /2$. This order parameter $\psi$ is essentially the sum of
the amplitude squares of all the modes. We have plotted the value of $\psi$
versus the chemical
potential
in fig. 4(b). It is
quite evident from the figure that there is a continuous transition at around
$\mu^{*}_{s}=4.8$.
{}From the above numerical results, we believe the morphology transition (in
our model) from the
DBM phase to the dendritic phase is continuous.

To further clarify our results,
we would like to comment on the simulations of ref.~\cite{a8} which claim to
show co-existence of the two distinct morphologies.
In that reference, the authors performed simulations with the
diffusion-transition scheme and
confined the growth to a channel with the channel direction oriented $45^o$
with respect to the preferred growth direction for possible dendrites. In the
parameter
range where one observes the dendrite phase in the open geometry, they obtained
a different morphology that is similar to DBM. They therefore claimed there
exists some range of parameter where both of
the two phases are ``available", and this coexistence is used to support the
idea that transition is discontinuous.
However, because of the
anisotropic nature of the dendritic phase, the boundary condition of the
channel
geometry in ~\cite{a8} has a very strong effect on the final pattern. It is
therefore impossible to identify the existence of coexisting phases by using
such argument.
One way to see this
is to note that if one continuously changes the angle between the
direction of the surface energy anisotropy and the boundary, one would
see a continuous change of the front velocity; this certainly should not
be attributed to the existence of a continuous family of
morphologies.

To explicitly show how the channel results can be misleading, we have studied
our mean field model eqs. (3,4) in the channel
geometry with
the lattice direction chosen to match the numerical experiment
of ref. ~\cite{a8}. We have used channel width $W=50$ and channel lengths
up to L=300. In the parameter regime where DBM exists, we see no change of
front velocity as shown in fig4(a). This is because DBM phase is a local
uncorrelated phase
which is not affected by the boundary. When we decrease the chemical potential
to the dendrite regime, the velocity of the front in the channel diverges
from that in the open geometry, going below the previous curve.
This is due to the
strong interaction between the channel wall and the dendrites which keep
hitting
the channel boundary and changing direction.
In fact, this observation is very useful in identifying the actual morphology
transition point; as we can see, the transition point determined by this
method is the same as determined by the order parameter $\psi$. On the other
hand, it should not be interpreted as coexistence of the two phases, as our
model has a continuous transition.

There is a standard and much more direct way to test whether a transition
is continuous. We can adiabatically change the parameter
across the transition point, and determine whether there is any hysteresis.
We prepare our system in the DBM regime very close to the transition
point $\mu_{s} =\mu^{*}_{s}$, when we slowly change the chemical potential
to lie below
$\mu^{*}_{s}$. The pattern immediately starts to generate dendrites at the
edge of the previously disordered DBM
pattern; these grow along the preferred direction, and the forming of the
dendrite takes over the entire dynamics. A series of plots showing this
transition are given in fig.5. The chemical potential $\mu_s$ is decreased from
$\mu_s=5.8$ (fig. 5(a))to $\mu_s=3.3$ (fig. 5(d)).
And, the reverse process is observed with increasing
chemical potential. We thus see no evidence for the existence of two
attractors with different basins of attraction.

\section{Summary}

We have proposed a mean field equation to describe the averaged
behavior of a class of discrete diffusion-limited growth model. Although
details of the microscopic kinetics will alter the detailed structure of
our model, there are universal features which are independent of these details.
We have shown that as long as there are balancing effects between growth and
melting, a 1D steady state growing solution (the ``saturated state") will
disappear at some critical point; instead, the 1D dynamics
can be described by the
spreading state, where the solid grows with the maximum density that
is bigger than the supply (the ``starved state"). In 2D, these two phases
become the DBM phase and the dendrite phase respectively. The DBM phase is
featureless and the density behind the front is uniform, whereas the dendritic
phase organizes itself into a correlated structure which then  allows higher
density in parts of the cluster than that could be globally
supplied by the liquid state.
{}From these properties of the two phases, we have constructed an order
parameter, which is the measure of uniformity of the solid cluster. The
behavior of the order parameter and non-existence of hysteresis
suggests that the transition is continuous.

Although we have discounted the purported numerical evidence for the
co-existence of the DBM and dendritic phases, there still remains the
discrepancy between our results on the velocity slope and those
reported in the simulation studies. Assuming that both our results and
these kinetic studies withstand further scrutiny, the only remaining
possibility is that one cannot ignore fluctuation effects even insofar
as determining the order of the transition. In any case, it is quite clear
that mean field approaches can only be used to obtain certain types of
qualitative
information - fluctuation effects are certainly important, probably
independent of spatial dimensionality, Including these in a complete
field-theoretic treatment of diffusion-limited growth remains a challenge
for future work.

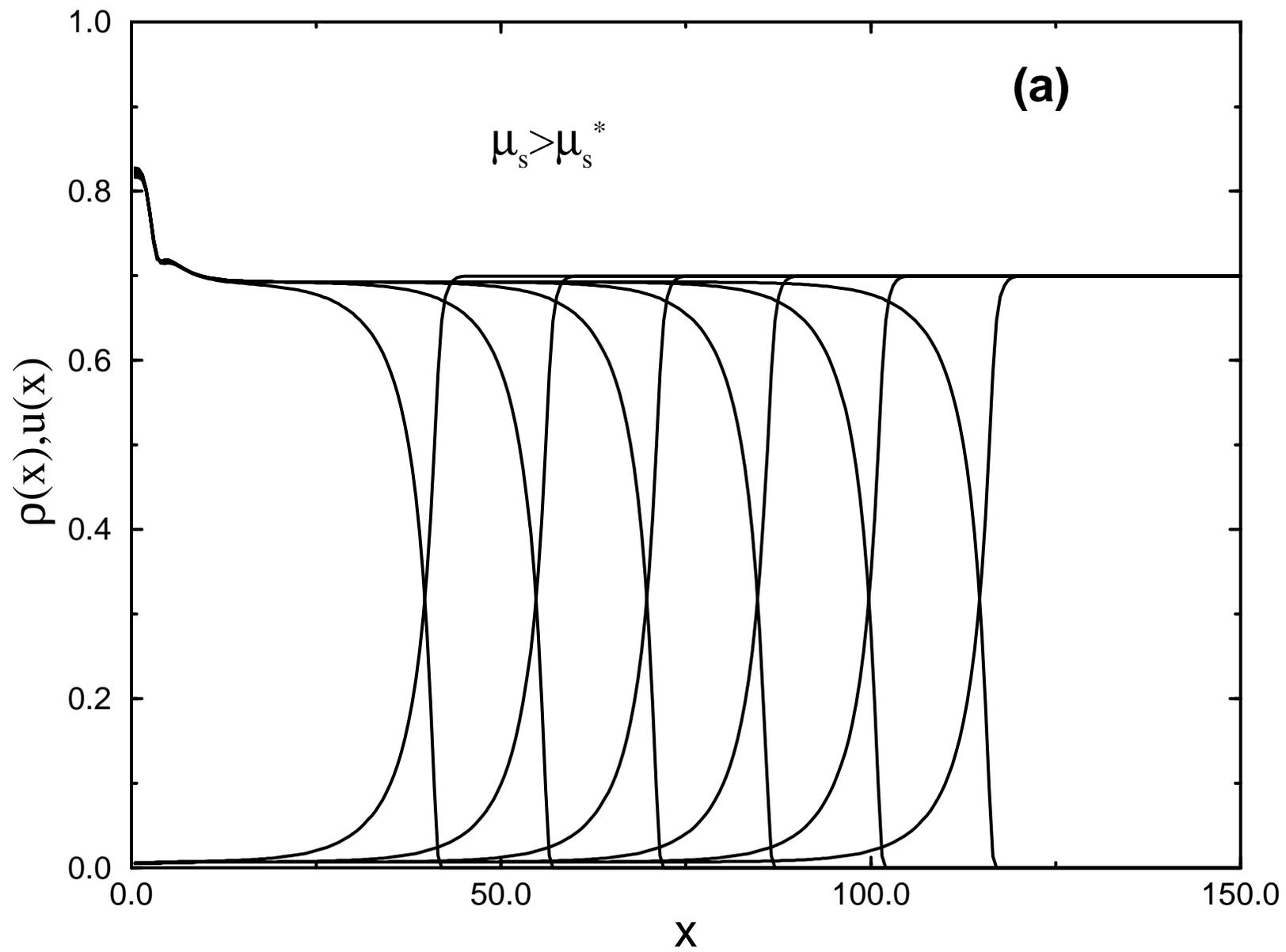
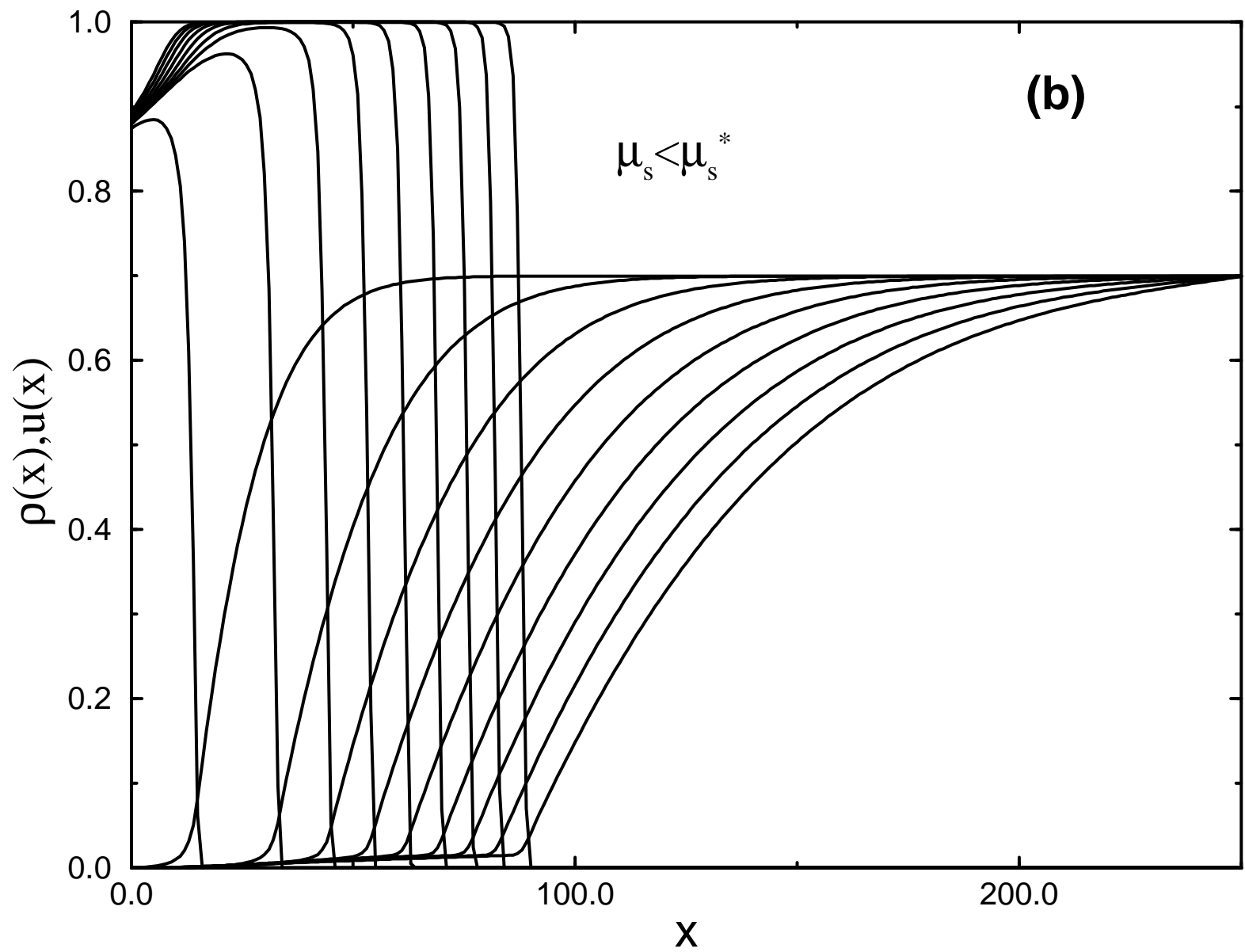
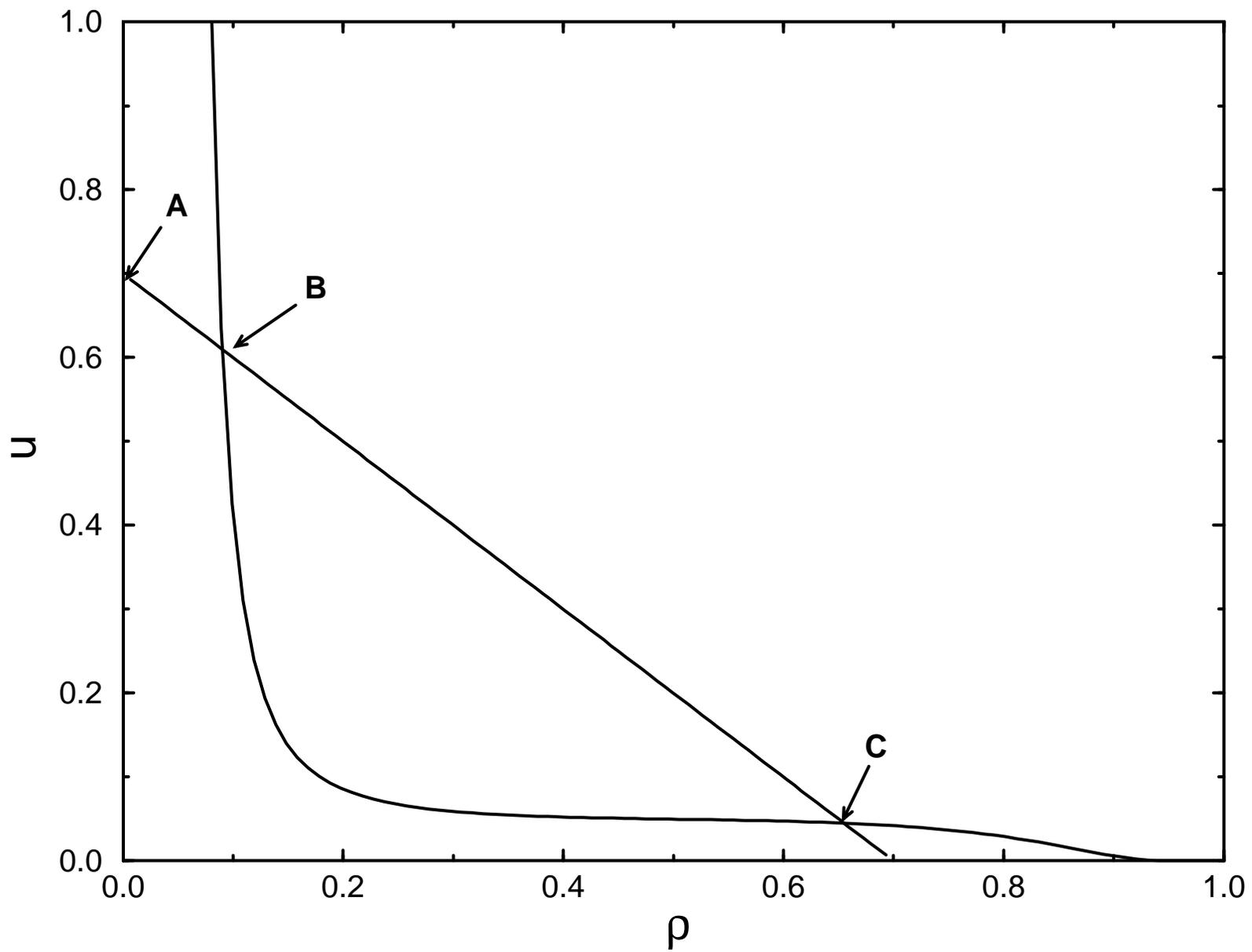
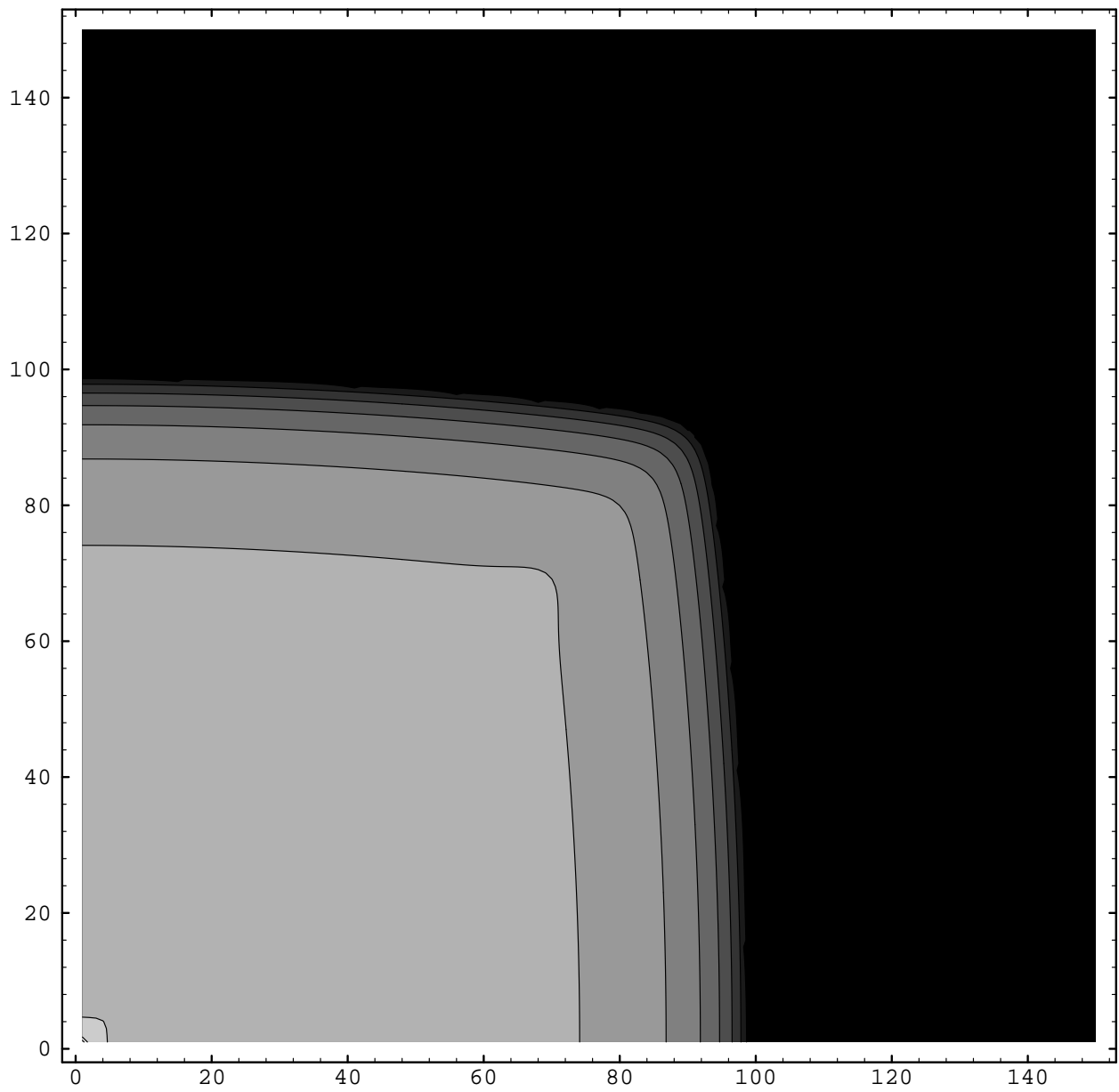
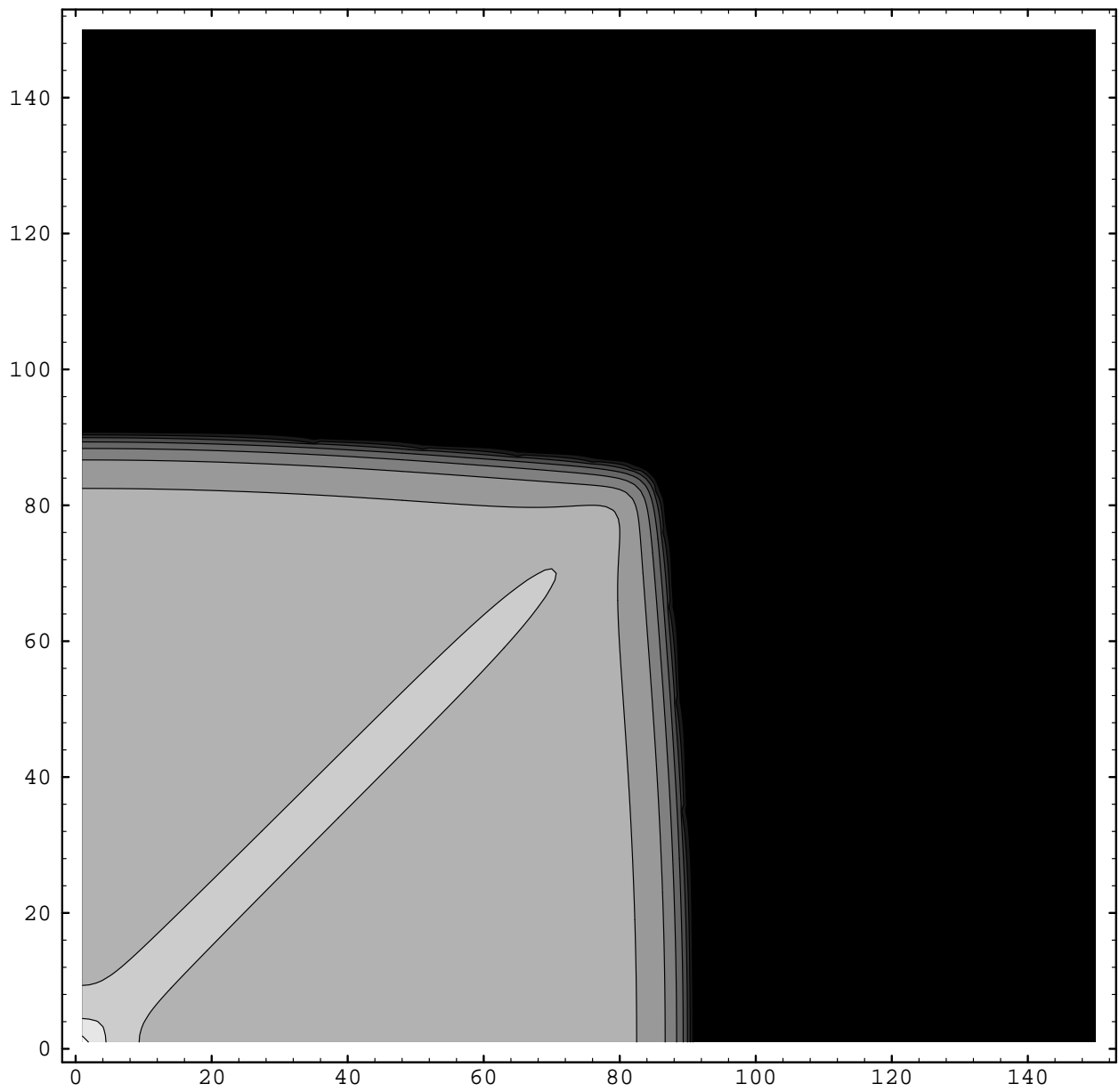
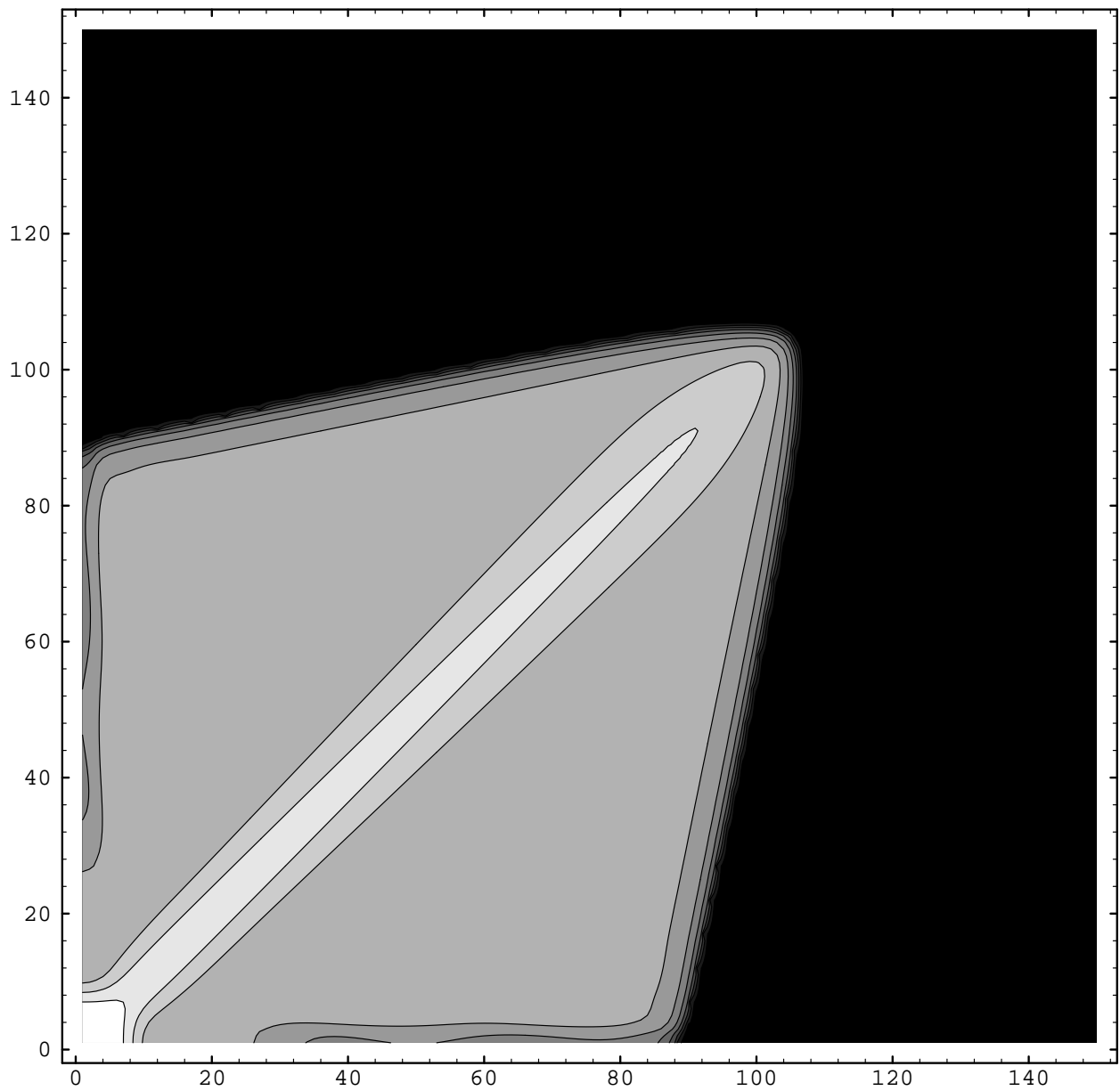
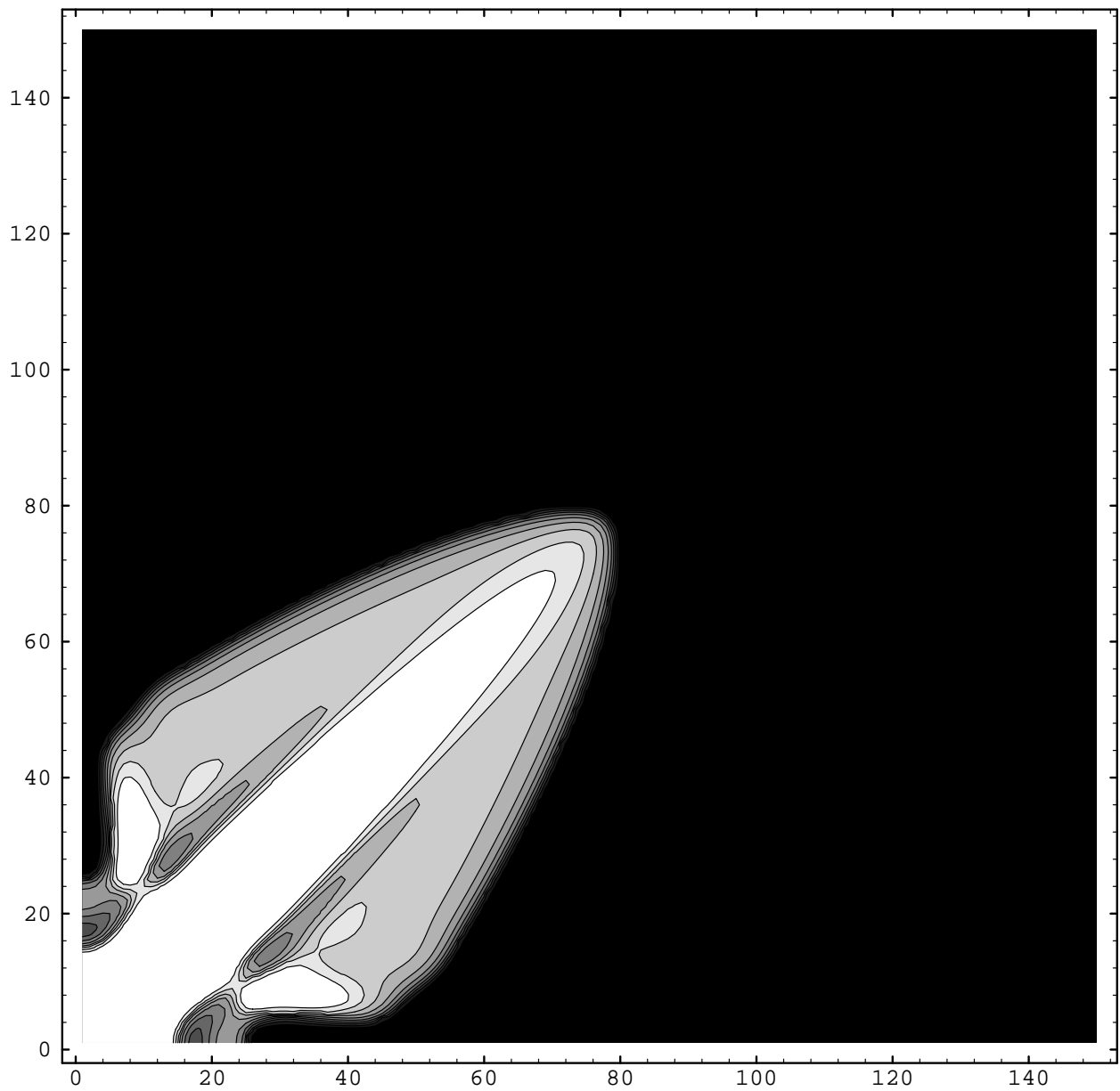
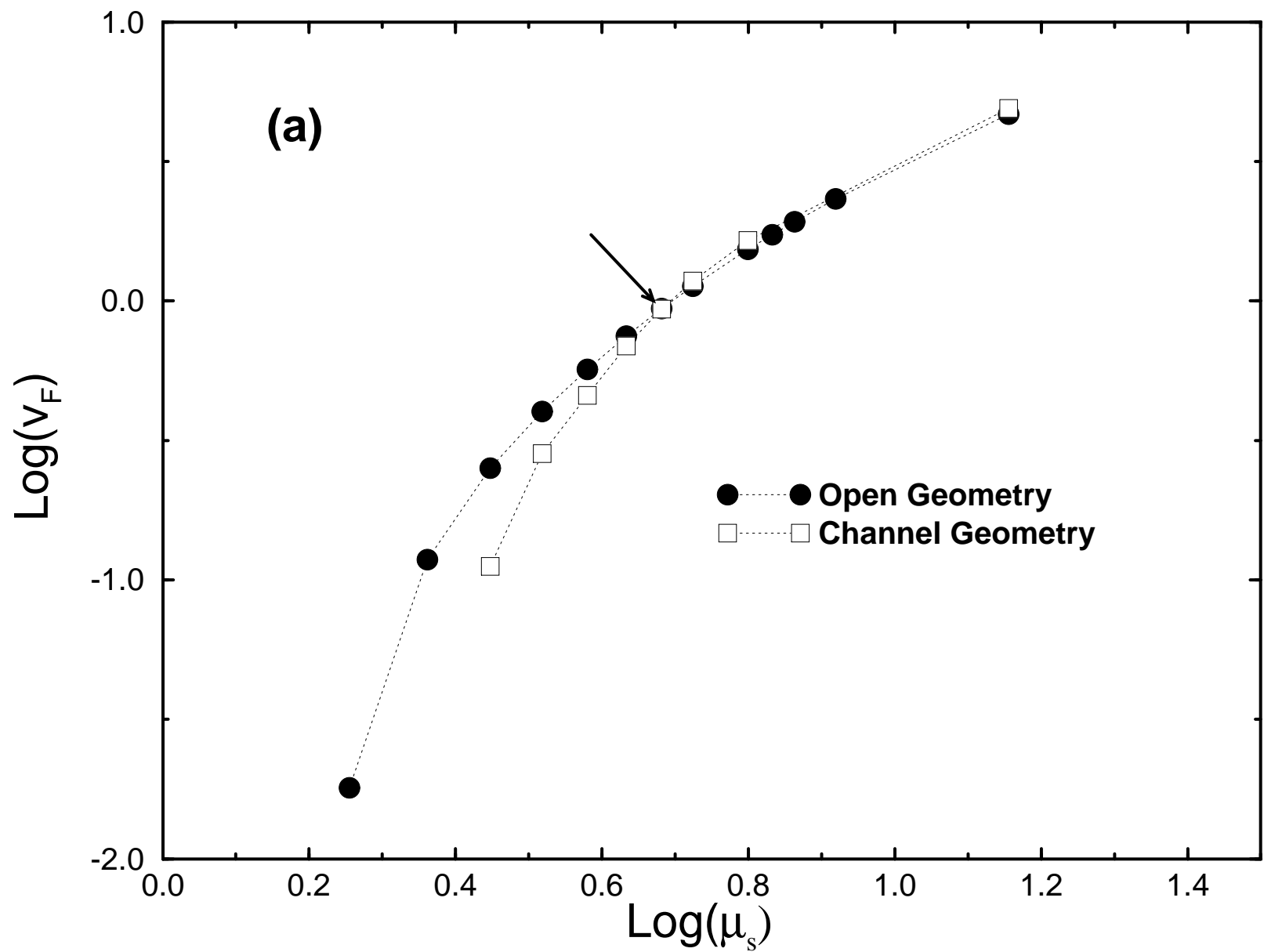
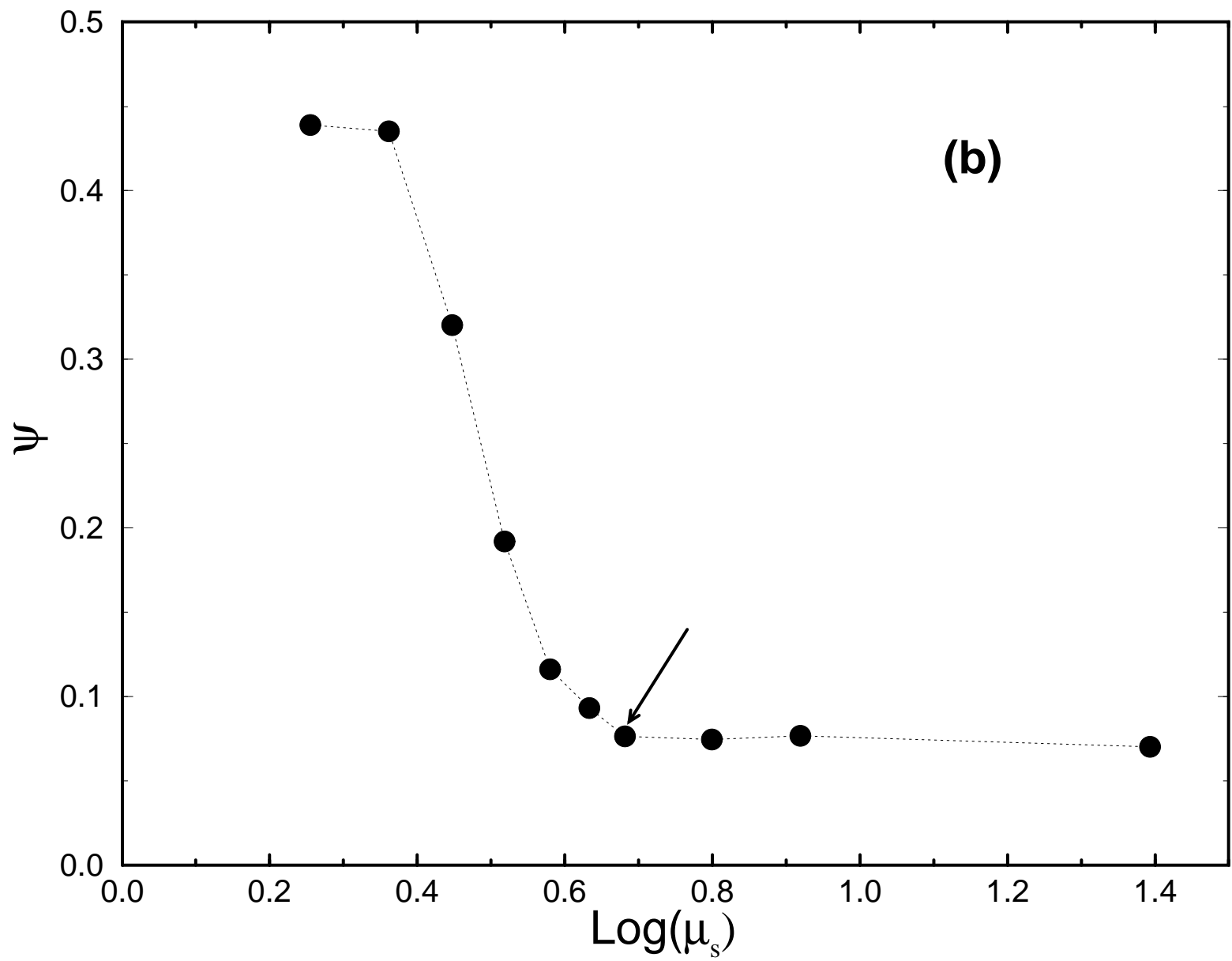
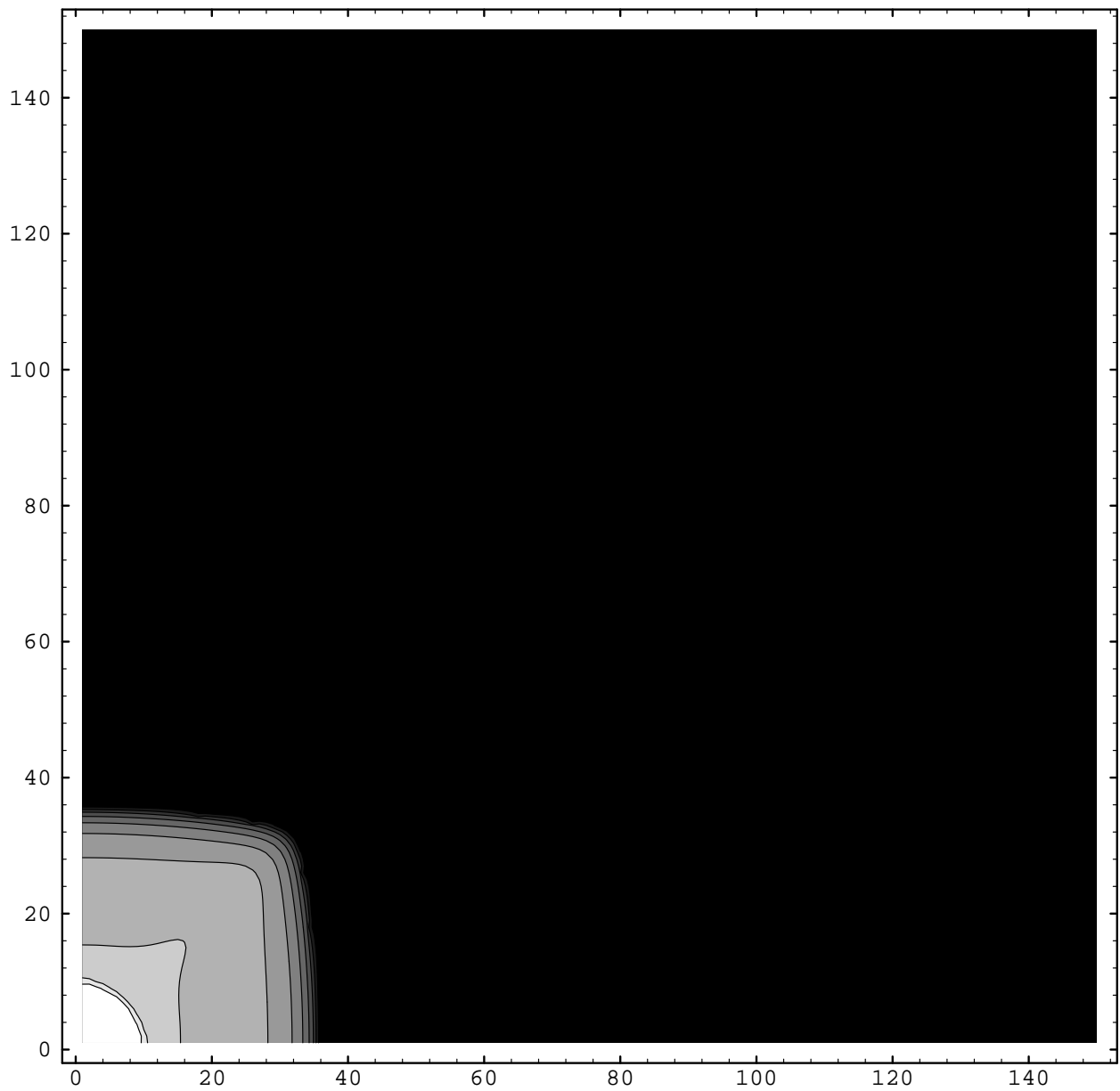
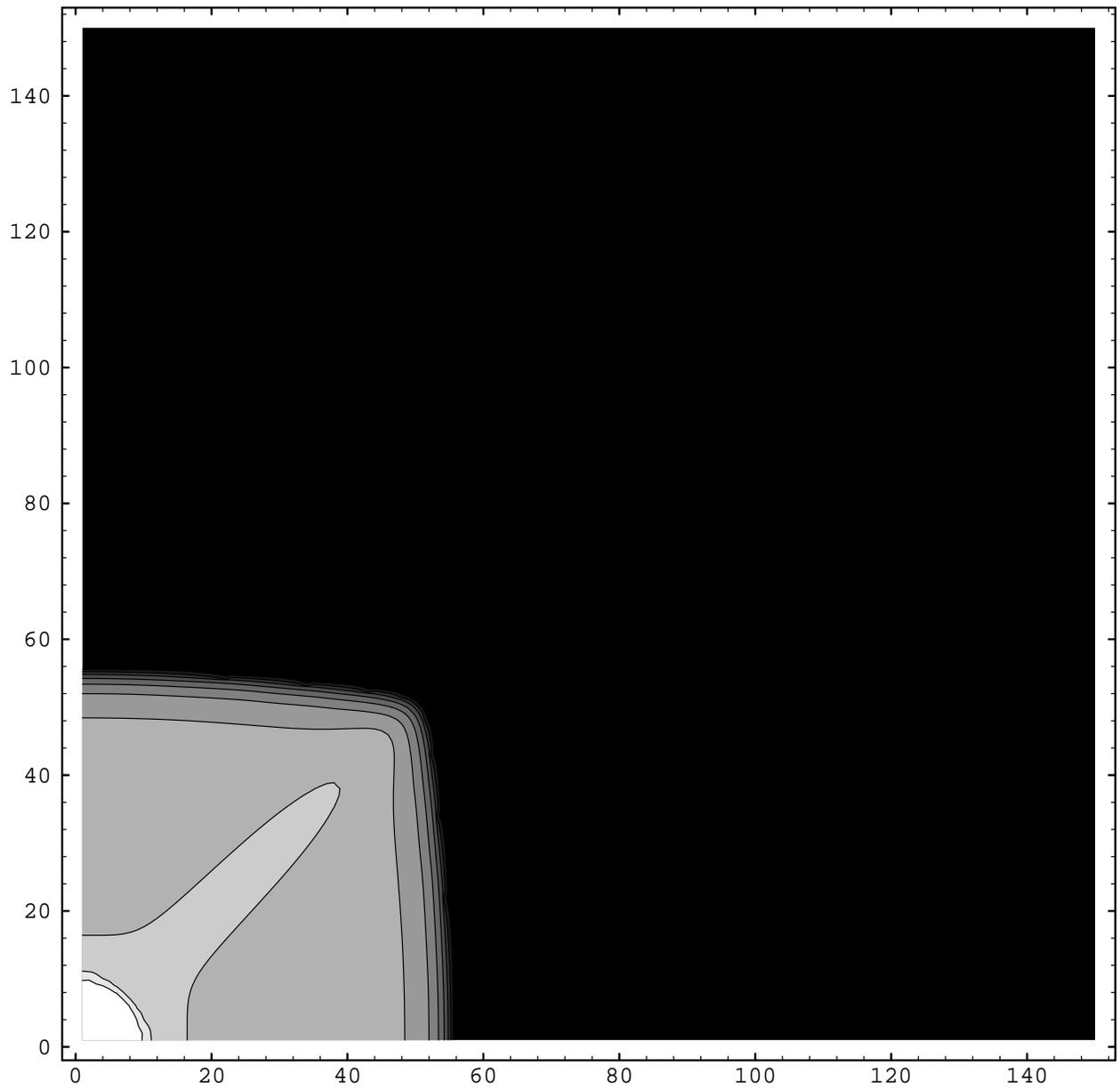
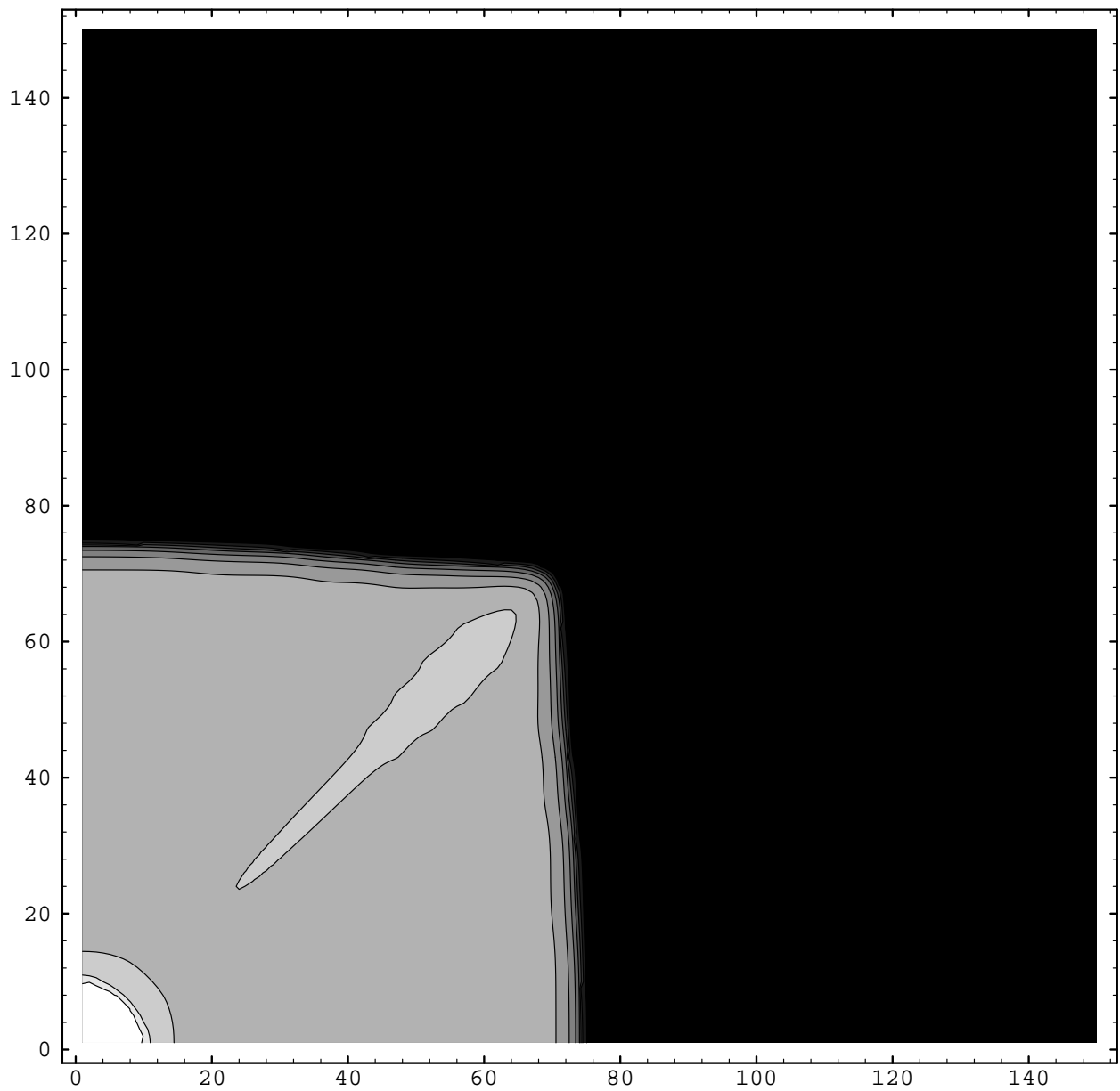
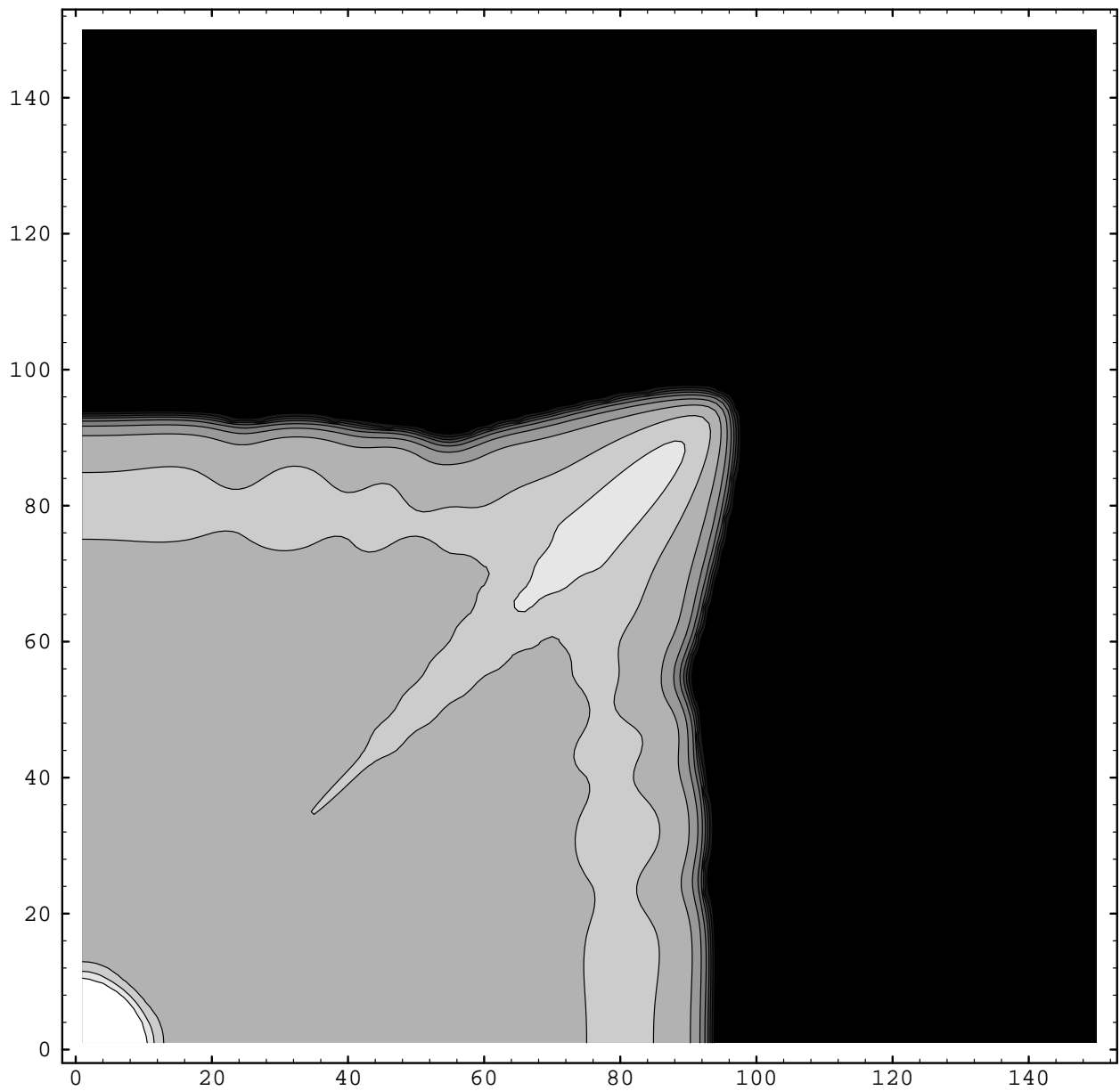
\end{document}